\documentclass[prd,twocolumn,nofootinbib,superscriptaddress]{revtex4}

  \usepackage[british]{babel}
  \usepackage{amsmath}
  \usepackage{amssymb}
  \usepackage{amsfonts}
  \usepackage{graphicx}

  \renewcommand{\eqref}[1]{Eq.~(\ref{#1})}
  
  \newcommand{\eg}{\emph{e.g.~}}

  \newcommand{\lab}{``lab'' }
  \newcommand{\V}[1]{\mbox{\boldmath $#1$}}
  \def\pc{{\rm pc}}

  \def\eV{{\rm eV}}
  \def\keV{{\rm keV}}
  \def\s{{\rm s}}

  \def\E{\V{E}}
  \def\B{\V{B}}
  \def\betaE{\V{\beta}_e}
  \def\betaP{\V{\beta}_p}
  \def\betaG{\V{\beta}_\gamma}
  \def\betaI{\V{\beta}_I}

  \def\thetaV{\V{\theta}^{(v)}}
  \def\thetaPi{\V{\theta}^{(\pi)}}
  \def\Vchi{\V{\chi}}

\begin{document}

\begin{minipage}[h]{\textwidth}
 \begin{flushright}t07/156\end{flushright}
\end{minipage}

\title{Challenges for creating magnetic fields by cosmic defects}

\author{Lukas Hollenstein}
\email{lukas.hollenstein@port.ac.uk} \affiliation{Institute of
Cosmology \& Gravitation,
             University of Portsmouth, Portsmouth PO1 2EG, UK}

\author{Chiara Caprini}
\email{chiara.caprini@cea.fr} \affiliation{Service de Physique Th\'eorique, CEA Saclay, F91191 Gif-sur-Yvette, France}

\author{Robert Crittenden}
\email{robert.crittenden@port.ac.uk} \affiliation{Institute of
Cosmology \& Gravitation,
             University of Portsmouth, Portsmouth PO1 2EG, UK}

\author{Roy Maartens}
\email{roy.maartens@port.ac.uk} \affiliation{Institute of
Cosmology \& Gravitation,
             University of Portsmouth, Portsmouth PO1 2EG, UK}

\date{\today}

\begin{abstract}

We analyse the possibility that topological defects can act as a source of magnetic fields through the Harrison mechanism in the radiation era. We give a detailed relativistic derivation of the Harrison mechanism at first order in cosmological perturbations, and show that it is only efficient for temperatures above $T\simeq 0.2\,\keV$. Our main result is that the vector metric perturbations generated by the defects cannot induce vorticity in the matter fluids at linear order, thereby excluding the production of currents and magnetic fields. We show that anisotropic stress in the matter fluids is required to source vorticity and magnetic fields. Our analysis is relevant for any mechanism whereby vorticity is meant to be transferred purely by gravitational interactions, and thus would also apply to dark matter or neutrinos.

\end{abstract}

\maketitle

\section{Introduction}

There is strong observational evidence for the existence of large-scale magnetic fields in galaxies and clusters, with intensity $0.1-10\,\mu{\rm G}$ (see, e.g., Refs.~\cite{mfobservational,mfreviews}). The observed features suggest a common generation mechanism, leading to weak seed fields that are amplified first by the collapse of structure and turbulent substructure formation, and then by the dynamo mechanism~\cite{Dynamo,Kulsrud:2007an}.
However, dynamo amplification is still under debate and depends not only on the type of structure but also on the cosmological parameters \cite{Kulsrud:2007an,Davis:1999bt}.

Magnetogenesis mechanisms can operate either primordially or during galaxy formation. Primordial magnetogenesis is particularly appealing, because it can produce seed fields on large scales, and would account for their ubiquity. Several primordial generation mechanisms have been proposed. For example, magnetic fields can be generated during a first order primordial phase transition~\cite{first}, and during the electroweak phase transition even if it is second order~\cite{second}; magnetic helicity can be generated by parity-violating processes~\cite{hel1,hel2}; magnetic fields can also be generated during inflation, if the conformal invariance of electromagnetism is broken in some way~\cite{inflation}, or during a pre-big bang~\cite{prebig}. These are only some examples of the generation mechanisms that have been proposed in the literature, many of which are strongly constrained by Nucleosynthesis~\cite{Caprini:2001nb}. It is fair to say that there is to date no preferred generation mechanism, and the origin of astrophysical magnetic fields remains an open problem.

In this paper, we concentrate on a mechanism first proposed by Harrison~\cite{Harrison:1970}. The basic idea is the following: during the early radiation era, Thomson scattering between electrons and photons is more efficient than Coulomb scattering between electrons and protons, so that electrons and photons can be described as a single fluid with a common velocity, in principle different from that of the protons. If, by means of some external mechanism, vortical motion is present in the radiative and proton fluids, then, in the absence of interactions, the vorticities of the two fluids evolve independently. The different spin-down rates of the photon-electron and proton fluids give rise to non-zero net rotational currents, which in turn can act as sources of magnetic fields. Two fundamental ingredients are needed for this generation process to work. Firstly, one needs an external source of vorticity in the fluids. Secondly, the proton and photon-electron fluids must have different velocities, otherwise there is no current and consequently no magnetic field.

In Harrison's original argument the analysis is Newtonian, and some of the assumptions are not made explicit. We present in section~\ref{sec:harrison} a detailed relativistic derivation of the Harrison mechanism, consistent at first order in cosmological perturbation theory. We show that the Harrison mechanism can operate when the temperature of the Universe is $T>0.2\,\keV$, so that Thomson scattering of electrons and photons is more efficient than Coulomb scattering of electrons and protons. Moreover, we show that the magnetic field sourced by the Harrison mechanism depends only on the initial value of the total vorticity in the fluids.

There is no obvious reason for vorticity to be non-zero, unless there is an external source. The vorticity might arise from second-order gravitational and scattering effects related to the nonlinear evolution of density perturbations, as studied in Refs.~\cite{Matarrese:2004kq,secondorderB,Kobayashi:2007wd,Takahashi:2007ds}. At first order, there is no vorticity from inflationary-generated perturbations, but topological defects are an active source of vector perturbations. In this paper we analyse the possibility that vorticity in the matter fluids could be driven by topological defects. We show that, at first order in perturbation theory, and in the absence of electrical currents in the defects themselves, this idea cannot work.

We assume that the defects interact with the cosmic fluid only gravitationally, and that they have no significant effect on the background evolution of the Universe. Defects are modelled as an inhomogeneously distributed component whose energy momentum tensor has scalar, vector and tensor degrees of freedom~\cite{defects_perts,Durrer:1998rw,Bevis:2006mj}. Therefore, defects induce perturbations in the metric at first order, and in particular vector-type perturbations~\cite{Durrer:gaugeinvariant}. Our main result is presented in section~\ref{sec:basics}, where we demonstrate that first-order vector-type metric perturbations cannot induce vorticity in perfect fluids without anisotropic stress. Consequently, defects cannot act as a source of magnetic fields in such fluids through the Harrison mechanism at first order. We identify fluid anisotropic stress as a possible way around this no-go result. (If the defects are super-conducting and carry primordial currents, then their magnetic fields can induce magnetic fields in the fluids, at first order~\cite{conductingstrings}.)

At early times, baryons and photons have negligible anisotropic stresses on the scales of interest, and consequently their vorticity is not sourced. On the other hand, defects have non-zero anisotropic stress, which induces vorticity in the defects and vector perturbations in the metric. However, we show that the vector modes in the defects and the metric form a closed system of conserved quantities and do not couple to the vorticity in the fluids.

Previous analyses have considered the possibility of Harrison magnetogenesis via vorticity from cosmic defects~\cite{strings1,Sicotte:1994ac,Avelino:1995pm,strings2,Battefeld:2007qn}. The argument relies on breaking the perfect fluid condition, and/or on nonlinear effects in the defect evolution. In Refs.~\cite{strings1}, the authors argue that for fluid particles that pass near enough to cosmic strings, shock fronts form in the fluid. The shock causes a discontinuity in the entropy of the fluid, and leads to turbulence and vorticity, and in turn to magnetic fields through a Harrison-type mechanism \cite{Mishustin:1971}. This effect, which is confined to the wakes after recombination and due to highly non-linear dynamics, is completely absent in our analysis, which restricts to linear theory and deals with perfect matter fluids with constant entropy. Subsequently it was proposed that magnetic fields can be generated also prior to recombination, around matter-radiation equality, via nonlinear dynamical friction of the surrounding particles on the motion of two wiggly strings moving in opposite directions~\cite{Avelino:1995pm,strings2}. In \cite{Battefeld:2007qn} it is argued that it is the rotating string loops rather than the long straight strings which dominate the generation of vorticity on the relevant scales.

In the next section we review the Harrison mechanism in first-order perturbation theory, and derive the collision rates for Coulomb and Thomson scattering. In section~\ref{sec:basics}, we present the necessary defect equations and then show that vorticity is not induced in the fluids if they do not have anisotropic stresses and do not interact other than gravitationally with the defects.

We consider a flat Fried\-mann-Le\-ma\^i\-tre-Ro\-bert\-son-Wal\-ker (FLRW) background. The metric perturbations can be split in scalar, vector, and tensor types, which decouple at linear order. Because we investigate the vortical dynamics of the system, we only consider vector-type metric perturbations:
\begin{equation}
  ds^2 \ = \ a^2(\tau)\left\{-d\tau^2 +2\chi_i\,d\tau dx^i +\delta_{ij}\,dx^i dx^j \right\} \,,
\end{equation}
where we choose the Poisson gauge, so the transverse three-vector $\Vchi$ is the only vector degree of freedom~\cite{kodama}. The scale factor $a$ is normalised so that $a(\tau_0)=1$ today.


\section{Harrison mechanism and magnetic field evolution}
\label{sec:harrison}

The Harrison mechanism in the radiation Universe is based on the mutual interaction of protons, electrons, photons (labelled by $I=p,e,\gamma$) and electromagnetic fields. The radiation and charged particles are modelled as perfect fluids with barotropic equation of state, $p_I=w_I\rho_I$, and with transverse velocity perturbation, $\V{v}_I$, relative to the comoving observer frame defined by the four-velocity $u^\mu=(a^{-1},0,0,0)$. Thus the four-velocities are:
\begin{equation}
  u_I^\mu \ = \ {a}^{-1}\left( 1, \V{v}_I \right) \quad \Rightarrow
  \quad u^I_\mu \ = \ a\left( - 1, \V{v}_I +\Vchi \right) \,,
\end{equation}
where $\V{k}\cdot \Vchi=0=\V{k} \cdot \V{v}_I$ (where $\V{k}$ is the comoving wave vector). Density perturbations $\delta\rho_I$ are of scalar type and contribute to the vorticity only at higher order, so that we do not consider them. The fluid vorticities are given by
\begin{equation}\label{vort}
  \omega^\mu_I \ = \ \frac{1}{2a}(0,\betaI)\,, \qquad
  \betaI \ := \ \frac{\rm i}{a}\, \V{k}\times(\V{v}_I+\Vchi) \,,
\end{equation}
as derived in Appendix~\ref{sec:vorticity}.

The fluid and electromagnetic energy-momentum tensors are
\begin{eqnarray}
  T^{\mu\nu}_I & = & (\rho_I+p_I)u_I^\mu u_I^\nu+p_I g^{\mu\nu} \,,
    \\
  T^{\mu\nu}_{\rm em} & = & \frac{1}{4\pi} \left[ (E^2+B^2)u^\mu u^\nu +\frac{1}{2}(E^2+B^2)g^{\mu\nu} \right.
    \nonumber \\
  &&\left. +2u^{(\mu}\varepsilon^{\nu)}_{\phantom{\nu}\gamma\delta} E^\gamma B^\delta-E^\mu E^\nu-B^\mu B^\nu\right] \,,
\end{eqnarray}
where $\varepsilon_{\mu\nu\gamma}= \eta_{\mu\nu\gamma\delta} u^\delta$ is the projected totally anti-symmetric tensor, and parentheses denote symmetrisation (see Appendix~\ref{sec:vorticity}). In the next section we will add cosmic defects to this system. Since the defects and the other components interact only gravitationally, the total energy-momentum tensor for the fluids and electromagnetic field is conserved:
\begin{eqnarray}
  && \nabla_\nu T^{\mu\nu}_{I} \ =\ \sum_JK_{JI}^\mu +K_{\text{em},I}^\mu \,,
    \quad \sum_{I,J}K_{JI}^\mu \ =\ 0 \,, \label{eq:sumT}
    \\
  && \nabla_\nu T^{\mu\nu}_{\text{em}} \ =\ -\sum_I K_{\text{em},I}^\mu \,.
\end{eqnarray}
The scattering and electromagnetic rates of momentum exchange are, at first order~\cite{Maartens:1998xg,Kobayashi:2007wd}
\begin{eqnarray}
  K^\mu_{IJ} & = & \frac{1}{a}\Big( 0, C_{IJ}[\V{v}_J-\V{v}_I] \Big) \,,
    \\
  K^\mu_{{\rm em},I} & = & \frac{1}{a}\Big( 0, q_In\E \Big) \,,
\end{eqnarray}
where $C_{IJ}$ are the Thomson or Coulomb collision coefficients, $q_I$ are the charges ($q_\gamma=0, q_p=e=-q_e$), $n=\rho_p/m_p=\rho_e/m_e$ is the number density, and $\E=\E(\tau,\V{k})$ is the transverse part ($\V{k} \cdot \E=0$) of the electric field as measured in the \lab frame. (See the appendices for further details. In \eqref{eq:KemI} of Appendix~\ref{sec:em}, we neglect the Lorentz force term which is second order.)

Using the above expressions for the momentum exchange rates, the fluid momentum conservation \eqref{eq:sumT} gives
\begin{eqnarray}
  && \Big[ (1+w_I)a^4\rho_I \left(\V{v}_I +\Vchi\right) \Big]'
    \nonumber \\
  && \hspace{1.5cm}= \ a^5\Big\{ \sum_J C_{IJ}(\V{v}_J-\V{v}_I) +q_I n \E \Big\} \,,
    \label{eq:FluidMomCons}
\end{eqnarray}
where a prime denotes a derivative with respect to conformal time $\tau$. Taking the curl, and using Maxwell's induction equation~(\ref{eq:MaxInductLab}),
\begin{equation}
  {\rm i}\V{k}\times(a^2\E) \ = \ - \left[a^2\B\right]'\,,  \label{eq:Induction}
\end{equation}
we arrive at the following system of equations:
\begin{eqnarray}
  \hspace{-1cm} &&\frac{4}{3}\left[ a^5\rho_\gamma \betaG \right]'
    \nonumber \\
  && \hspace{1cm} = a^6\left\{ C_{e\gamma}(\betaE-\betaG) + C_{p\gamma}(\betaP-\betaG) \right\},
    \label{eq:PhotonVort} \\
  \hspace{-1cm} &&\left[ a^5\rho_e\betaE \right]' -ena^3\left[a^2\B\right]'
    \nonumber \\
  && \hspace{1cm} = a^6\left\{ C_{e\gamma}(\betaG-\betaE) + C_{ep}(\betaP-\betaE) \right\},
    \label{eq:ElectronVort} \\
  \hspace{-1cm} && \left[ a^5\rho_p\betaP \right]' +e na^3\left[a^2\B\right]'
  \nonumber \\
  && \hspace{1cm} = a^6\left\{ C_{p\gamma}(\betaG-\betaP) + C_{ep}(\betaE-\betaP) \right\}.
    \label{eq:ProtonVort}
\end{eqnarray}
These equations relate the vorticity in the three fluids (photons, electrons and protons) to the magnetic field. In order to close the system, we use the curl of Amp\`ere's law \eqref{eq:CurlAmpereLab} which leads to a wave equation for the magnetic field sourced by the vortical current,
\begin{equation}
  \left[a^2\B\right]'' +k^2a^2\B \ = \ {4\pi en} a^4\, (\betaP-\betaE) \,,  \label{eq:Bwave}
\end{equation}
such that Eqs.~(\ref{eq:PhotonVort})--(\ref{eq:Bwave}) describe the generation and evolution of magnetic fields in the radiation era at first order (the electric and magnetic fields vanish in the background).

We now investigate the collision terms on the right hand sides of Eqs.~(\ref{eq:PhotonVort})--(\ref{eq:ProtonVort}), which determine the momentum exchange among the particle species. The coefficients $C_{IJ}$ are the ratios between the enthalpy density of the scattering particles and the mean relaxation times for scattering between the species~\cite{Harrison:1973bt,Kobayashi:2007wd}:
\begin{eqnarray}
  C_{ep} & := & \frac{\rho_e}{\tau_{ep}} \ = \ \frac{\rho_p}{\tau_{pe}}
    \ = \ \frac{4\pi e^4\ln\Lambda}{m_e}\,n^2 \Big(\frac{m_e}{T}\Big)^{3/2},
    \\
  C_{e\gamma} & := & \frac{\rho_e}{\tau_{e\gamma}}\ =\ \frac{4\rho_\gamma}{3\tau_{\gamma e}}\ =\
    \frac{32\pi e^4 }{9m_e^2}\,\rho_\gamma n \,,
    \\
  C_{p\gamma} & := & \frac{\rho_p}{\tau_{p\gamma}}\ =\ \frac{4\rho_\gamma}{3 \tau_{\gamma p} }\ =\
    \Big(\frac{m_e}{m_p}\Big)^2\,C_{e\gamma} \,,
\end{eqnarray}
where $\ln\Lambda\simeq 17$ is the Coulomb logarithm, coming from the Coulomb cross-section, $\sigma_C=4\pi e^4\ln\Lambda/T^2$. In the radiation era the relaxation times are
\begin{eqnarray}
  \tau_{ep} & := & \frac{1}{v_e n\sigma_C} \ \simeq \ 9\times
    10^{-4} \Big(\frac{\eV}{T}\Big)^{3/2} \,\s,
    \label{eq:tauep} \\
  \tau_{e\gamma} & := & \frac{3m_e}{4\rho_\gamma\sigma_T} \ \simeq \
    7\times 10^2 \Big(\frac{\eV}{T}\Big)^{4} \,\s,
    \label{eq:tauegamma} \\
  \tau_{p\gamma} & := & \frac{3m_p}{4\rho_\gamma\sigma_{Tp}} \ = \
    \Big(\frac{m_p}{m_e}\Big)^3\,\tau_{e\gamma} \,,
    \label{eq:taupgamma}
\end{eqnarray}
where $\tau_{IJ}$ denotes the scattering of the $I$-type particle against the $J$-type particle, $v_e$ is the thermal electron velocity, $\sigma_T$ is the electron-photon Thomson cross-section and $\sigma_{Tp}=(m_e/m_p)^2\sigma_T$ is the Thomson cross-section of proton-photon interactions. Here and subsequently we use values for the cosmological parameters from \cite{Lahav:2006qy}.

\begin{figure}[ht]
  \begin{center}
    \includegraphics[width=\columnwidth]{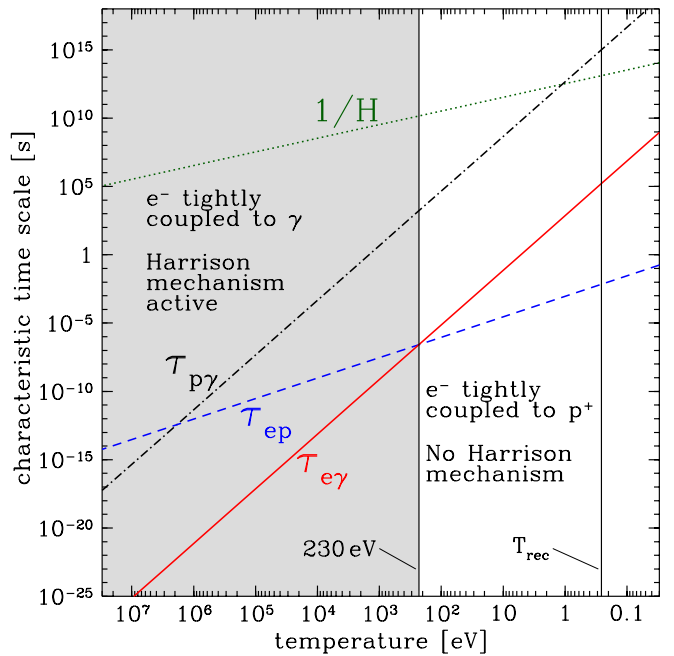}
      \caption{The mean relaxation times, Eqs.~(\ref{eq:tauep})--(\ref{eq:taupgamma}), and the expansion time-scale, $1/H$, as a function of the radiation temperature. The Harrison mechanism operates in the shaded region, $T\gtrsim T_c\simeq 230\,\eV$.}
    \label{fig:TimeScales}
  \end{center}
\end{figure}

The evolution of the various fluid velocities and vorticities is governed by the relative strengths of these relaxation times, which are shown in Fig.~\ref{fig:TimeScales}. At high temperatures, the strongest coupling is between the electrons and the photons, and the scattering rate is so high that they can be treated as tightly coupled.  At lower temperatures, the photon density is diluted and eventually the electron-proton interactions come to dominate. This occurs when $C_{ep}\simeq C_{e\gamma}$, at a cross-over temperature of $T_c\simeq 230\,\eV$. A similar transition occurs between the proton-photon scattering and the proton-electron scattering at a higher temperature, $T\simeq 94\,\keV$. For temperatures $T\lesssim 94\,\keV$ we can neglect the proton-photon coupling, and we make this approximation from now on.

Recently Takahashi et al.~\cite{Takahashi:2007ds} have pointed out that electrons and protons are coupled not only by Coulomb scattering, but also via their coupling to the electric field, and they find that the latter coupling is much more efficient than Coulomb scattering. This would mean that Thomson scattering is dominant only at higher temperatures than $\simeq 230\,\eV$, i.e., $T_c$ would be higher than shown in Fig.~1.

At temperatures below $T_c$, the electron-proton coupling dominates, which damps any currents that are needed for magnetogenesis. A difference in the vorticities of the electron and proton fluids is in fact necessary for magnetic field generation, as appears clearly from equation (\ref{eq:Bwave}).

Magnetogenesis becomes possible when photon pair production ceases and the three-fluid picture described above holds. We assume there is no magnetic field initially, $\B(a_{\rm i})=0$, so that due to \eqref{eq:Bwave} and the tight coupling of all fluids the vorticities are equal, $\betaG{}_{\rm i}=\betaE{}_{\rm i}=\betaP{}_{\rm i}=:\V{\beta}_{\rm i}$. We can derive a total vorticity evolution equation by adding all the vorticity exchange equations (\ref{eq:PhotonVort})--(\ref{eq:ProtonVort}). The exchange terms cancel and we arrive at the conservation of total angular momentum
\begin{equation}
  \frac{m_e}{m_H}a^2\betaE + \frac{m_p}{m_H}a^2\betaP +a_{\rm eq}a \betaG
    = a_{\rm i}^2 \left(\!1+\frac{a_{\rm eq}}{a_{\rm i}}\!\right) \V{\beta}_{\rm i} \,.
    \label{eq:TotFluidVortCons}
\end{equation}
Here we have defined $a_{\rm{eq}} ~(\simeq 1/680)$ via $\rho_b(a_{\rm{eq}}) = 4\rho_\gamma(a_{\rm{eq}})/3$. While this equation holds, vorticity is not created, but can be passed between the various species.

We derive the magnetic field evolution equation for the high temperature regime ($T\geq T_c$) where the photons and the electrons are tightly coupled. The Thomson collision term of the photons and electrons diverges in the tight coupling limit and must therefore be cancelled by adding Eqs.~(\ref{eq:PhotonVort}) and (\ref{eq:ElectronVort}). (Note that it is incorrect to simply drop the Thomson scattering terms.) After combining Eqs.~(\ref{eq:PhotonVort}) and (\ref{eq:ElectronVort}) we may safely set $\betaE=\betaG$, to find
\begin{equation}
  \left[ a^5\left(\!\frac{4}{3}\rho_\gamma +\rho_e\!\right)\betaG \right]' -ena^3\left[a^2\B\right]'
    = a^6C_{ep}(\betaP-\betaG)  \,. \label{eq:PhotElecVort}
\end{equation}
Then we can derive an evolution equation for the magnetic field by subtracting the proton vorticity equation (\ref{eq:ProtonVort}) from \eqref{eq:PhotElecVort}, and by using angular momentum conservation, \eqref{eq:TotFluidVortCons}, and the wave equation (\ref{eq:Bwave}), both with $\betaE=\betaG$. The result is given in Appendix \ref{sec:em}, \eqref{eq:MFEvol}. For sub-horizon scales, $k\gg aH$, it can be integrated to give
\begin{eqnarray}
  && \left(1+k^2\ell^2a\right)\B \ + \ \frac{L^2k^2}{a^2} \int_{a_{\rm i}}^a a^{5/2}\B\, {\rm d}a
    \nonumber \\
  &&~~~~~~{} \ = \ \frac{m_p}{e} \frac{a_{\rm i}}{a}\left(\frac{a_{\rm i}}{a}-1\right)\V{\beta}_{\rm i}\,.
    \label{eq:Bfull}
\end{eqnarray}
Here we used the fact that $a\ll a_{\rm eq}$ for $T\geq T_c$ and introduced the diffusion scales:
\begin{equation}
  L^2 \ :=\ \frac{e^2\sqrt{m_e}\ln\Lambda}{T_0^{3/2}H_0\sqrt{\Omega_{\rm rad}}}
    \,, ~~~~~{} \ell^2\ :=\ \frac{2m_Hm_pG}{3e^2H_0^2\Omega_{b}}\,.
\end{equation}
$L$ is a characteristic scale for Coulomb diffusion while $\ell$ is related to the classical electromagnetic interaction between the charged particles. Clearly the magnetic field is only sourced by a non-zero initial vorticity $\V{\beta}_{\rm i}$.

Equation~(\ref{eq:Bfull}) is the generalised form of Harrison's result, including the scale-dependent Coulomb diffusion integral term. We arrive at Harrison's result~\cite{Harrison:1970} by considering cosmologically relevant scales: well inside the horizon, but large enough that the diffusion terms are negligible, $k^{-1} \gg L\simeq 3.2\times 10^{-2}\,\pc \gg \ell \simeq 3\times 10^{-9}\,\pc$:
\begin{equation}
  \frac{e}{m_p}\B \ = \ \frac{a_{\rm i}}{a}\left(\frac{a_{\rm i}}{a}-1\right)\V{\beta}_{\rm i} \,.
    \label{eq:Bharrison}
\end{equation}

From Eqs.~(\ref{eq:Bfull}) and (\ref{eq:Bharrison}), it is clear that one needs initial vorticity to create magnetic fields. If the initial vorticity is zero, then under the assumption $\betaE=\betaG$, Eqs.~(\ref{eq:PhotElecVort}) and (\ref{eq:Bfull}) show that the vorticities and the magnetic field remain zero.

\section{Vorticity from defects?}
\label{sec:basics}

In order to produce a magnetic seed by Harrison's mechanism, the initial value of the total vorticity in the electron, proton and photon fluids must be non-zero. However, this begs the question of where this vorticity originated.  Harrison had turbulence in mind, but it now seems unlikely that significant turbulence will exist in the early Universe on the scales of interest~\cite{Rees:1987}. Vorticity originating in the very early Universe, such as during an inflationary period, would have decayed to negligible levels before the onset of Harrison's mechanism. However, cosmic defects are an active source of vector perturbations, at the same order of magnitude as scalar perturbations, which in principle could lead to a seed for magnetic fields.

Cosmic defects are usually modelled as a fluid~\cite{defects_perts}. Here we only consider their vector contribution, and the vector part of the defect energy-momentum tensor is
\begin{eqnarray}
  \Theta_{0i} & = & \theta^{(v)}_i
    \\
  \Theta_{ij} & = & \frac{\rm i}{2}\left[ k_i\theta^{(\pi)}_j +k_j\theta^{(\pi)}_i \right] \,.
\end{eqnarray}
Here $\theta^{(v)}_i$ is the transverse momentum density, and $\theta^{(\pi)}_i$ is the transverse vector part of the anisotropic stress. The \lab{}frame quantities are defined via the four-vectors $\theta^{(\alpha)}_\mu=a(0,\V{\theta}^{(\alpha)})$ for $\alpha=v,\pi$.

Because the defects interact only gravitationally with the radiation and matter fluids, they are separately conserved: $\nabla_\nu \Theta^{\mu\nu}=0$. Momentum conservation for the defects implies
\begin{eqnarray}
  \left[ a^3\thetaV \right]' \ = \ -\frac{1}{2}a^3k^2\thetaPi \,.  \label{eq:DefectsMomentum}
\end{eqnarray}
The total momentum conservation for the fluids follows from \eqref{eq:FluidMomCons},
\begin{equation}
\left[ a^4\sum_I (1+w_I)\rho_I(\V{v}_I+\Vchi) \right]' \ = \ 0 \,. \label{eq:TotFluidMomCons}
\end{equation}
Anisotropic stress in the defects sources the vector metric perturbation via the Einstein equation,
\begin{eqnarray}
  \left[ a^2\Vchi \right]' \ = \ 8\pi G a^3\thetaPi \,.  \label{eq:Dynamical}
\end{eqnarray}
However, this is not enough to source vorticity in the radiation and matter fluids, which remains conserved by Eq.~(\ref{eq:TotFluidVortCons}). The vorticity is not simply the curl of the fluid velocity -- by Eq.~(\ref{vort}) it is the curl of $\V{v}_I+\Vchi$. Defects source $\Vchi$, but cannot break the conservation of angular momentum. 

Therefore, we can conclude that {\em despite the fact that the defects change the evolution of the metric, the fluid velocities adjust to compensate this change and ensure the conservation of the vorticity.} Thus the defects do not induce magnetic fields via the Harrison mechanism at first order.

Note that the Einstein constraint equation does not provide any additional avenue for vorticity generation. It takes the form of a vector ``Poisson'' equation,
\begin{equation}
  a^2k^2\Vchi \ = \ -16\pi G \V{Q} \,,
\end{equation}
where we defined the total vector momentum (fluids and defects)
\begin{equation}
  \V{Q} \ := \ a^3\thetaV + a^4\sum_I(1+w_I)\rho_I \left( \V{v}_I +\Vchi \right) \,.
\end{equation}
From the dynamical Einstein equation (\ref{eq:Dynamical}) together with the momentum conservation equations (\ref{eq:DefectsMomentum}) and (\ref{eq:TotFluidMomCons}) we find
\begin{equation}
  \left[ a^2 k^2 \Vchi \right]' \ = \ -16\pi G \V{Q}'\,.
\end{equation}
Thus the constraint is identically satisfied at all times if it is satisfied at an initial time $\tau_{\rm i}$. This determines the initial metric vector perturbation in terms of the initial total momentum.

We have given a comprehensive general relativistic analysis which shows how vorticity is related to metric vector perturbations and to vector anisotropic stress. This analysis has not used any special properties of the defects. In fact, it appears that it is not possible to induce vorticity in the matter fluids by any source that interacts with the fluids purely gravitationally. Thus our general analysis would apply equally well to neutrinos (after decoupling) or dark matter. This result is implicit in Ref.~\cite{Battefeld:2007qn} for the special case of cosmic string wakes in a Newtonian approximation: in this reference angular momentum conservation is identified as the reason for the absence of vorticity at first order.

One possible source of vorticity in the fluid is {\em fluid} vector anisotropic stress \cite{Rebhan:1991sr}:
\begin{equation}
  \Pi^I_{ij} \ = \ \frac{{\rm i}p_I}{2}\left( k_i\pi^I_j + k_j\pi^I_i \right) \,,
\end{equation}
where $\pi^I_i$ is transverse and the \lab{}frame quantities are defined via the four-vectors $\pi^I_\mu=a(0,\V{\pi}_I)$.

Our key underlying assumption above is that none of the fluids (electrons, protons, photons) has significant anisotropic stress in the relevant temperature range $T \ge T_c$, since they are too tightly coupled with each other. If present, fluid anisotropic stress would not only provide an extra source for the metric perturbation,
\begin{eqnarray}
  \left[ a^2\Vchi \right]' \ = \ 8\pi G a^3\left\{\thetaPi +\sum_I p_I\V{\pi}_I \right\} \,,
    \label{eq:DynamicalFull}
\end{eqnarray}
but it would also directly source vorticity in the fluids via the curl of the momentum equation:
\begin{eqnarray}
  \left[ a^5\sum_I(1+w_I)\rho_I \betaI \right]' \ = \ -\frac{{\rm i}k^2a^3}{2}\V{k}\times\sum_I p_I \V{\pi}_I \,.
    \label{eq:TotalFluidMomentum}
\end{eqnarray}
This shows explicitly how fluid anisotropic stress violates the vorticity conservation shown above in \eqref{eq:TotFluidVortCons}. For fluids that support anisotropic stress it is possible that this can be sourced by the defects: \eg{} during recombination, when photons decouple from the baryons, metric vector perturbations from defects are a source for the photon anisotropic stress \cite{Durrer:1998rw}.

Another possible way to break our no-go result is at higher order in perturbations. For example, Kobayashi et al.~\cite{Kobayashi:2007wd} showed recently that at second order in perturbation theory and second order in the tight coupling approximation, vorticity and magnetic fields could be generated during the phase dominated by electron-proton scattering.

\section{Summary and conclusions}

In this paper we investigated the possibility that the seeds for the magnetic fields observed today in galaxies and clusters are generated in the primordial Universe by the Harrison mechanism. We found that the Harrison mechanism can act during a relatively early stage of the radiation era, for $T\geq T_c \simeq 230\,\eV$. At lower temperatures the electrons are tightly coupled to the protons and no current can be generated. We analysed the problem at first order in perturbation theory, and generalised Harrison's original result, showing that initial vorticity in the matter and photon fluid sources the magnetic field.

The idea of sourcing vorticity in the matter fluids via topological defects (in particular, cosmic strings) has been considered previously in the literature, based on Newtonian analyses of non-linear processes in the dynamics of one or more cosmic strings, such as turbulence induced in the wake of a string after recombination~\cite{strings1,Avelino:1995pm}. We analysed the problem at first order in gravitational effects, modelling the defects as a fluid with anisotropic stress. We found that the defects do not provide a source for vorticity in the matter fluids, so that no magnetic field is generated in the period of the radiation era relevant for the Harrison mechanism.

The reason for this is connected to the nature of vector perturbations: the fluid velocities adjust to compensate the metric perturbation sourced by defects, in such a way that fluid vorticity remains conserved. This result is not specific to defects: in a two-component system interacting only through gravity, the vector metric perturbations induced by one component do not act as a source of vorticity in the other component. At first order, the only source of vorticity is vector anisotropic stress in the fluid itself. If this is absent, no vorticity can be generated in the fluid even though the vector metric perturbation is non-zero.

For the Harrison mechanism to work, one requires either fluid anisotropic stress, or second-order effects. There is an interesting tension between the creation of vorticity and the creation of magnetic fields. The creation of vorticity requires anisotropic stress which is suppressed by strong coupling between the fluids. On the other hand, creating magnetic fields by the Harrison mechanism relies on strong coupling.

\acknowledgments

The authors thank Konstantinos Dimopoulos, Ruth Durrer, J\"urg Fr\"ohlich, Rajesh Gopal, Tsutomu Kobayashi, Martin Kunz, Keitaro Takahashi and Daniel Wyler for helpful discussions. The work of RC and RM is supported by STFC. CC acknowledges support from the ANR funding PHYS@COL\&COS.

\appendix

\section{Three-vectors, curls, and vorticity} \label{sec:vorticity}

For a flat FLRW metric with first-order scalar, vector and tensor perturbations in the Poisson gauge
\begin{eqnarray}
  ds^2 & = & a^2(\tau)\Big\{ -(1 +2\psi)\,d\tau^2 + 2\chi_i\,d\tau dx^i
    \nonumber \\
  && \phantom{a^2(\tau)\Big\{} +\left[ (1 -2\phi)\delta_{ij}+ h_{ij} \right]  dx^i dx^j \Big\} \,.
\end{eqnarray}
The observer at rest in the background (so that $\tau=\,$const are global rest spaces) has four-velocity
\begin{equation}
  u^\mu \ = \ \frac{dx^\mu}{dt}\ =\ \frac{1}{a}\frac{dx^\mu}{d\tau}\ =\ \frac{1}{a}\Big(1,0,0,0\Big) \,.
\end{equation}
The vector and tensor metric perturbations are defined as measured in FLRW proper co-ordinates $(t,\V{r})$, where $dt=ad\tau$, $d\V{r}=ad\V{x}$. We refer to this as the \lab frame (see Ref.~\cite{Subramanian:1997gi}), and in this frame there is no difference between upper and lower spatial indices (e.g., $\chi_i=\chi^i$).

For any four-vector $Z^\mu$ that is orthogonal to $u^\mu$, i.e., $u_\mu Z^\mu=0$, it is convenient to define the associated three-vector $\V{z}$ as measured in the \lab{}frame:
\begin{equation}
  Z^\mu\ =\ a^{-1}(0,\V{z}) \quad \Rightarrow \quad Z_\mu \ = \ a(0,\V{z}) \,.
    \label{eq:orthog}
\end{equation}
Then one can apply the usual Euclidean vector calculus to $\V{z}$.

With four-dimensional perturbed vector and tensor quantities one has to be careful, because raising and lowering indices of four-vectors is done with the perturbed metric. In the case of the four-velocity of fluid $I$,
\begin{equation}
  u_I^\mu \ = \ \frac{1}{a}\Big( 1-\psi, \V{v}_I \Big) \,, \quad
    u^I_\mu \ = \ a\Big( -1-\psi, \V{v}_I +\Vchi \Big) \,,
\end{equation}
the vector perturbation $\Vchi$ appears in the co-vector $u_\mu^I$ because $u_\mu^I$ is not orthogonal to the observer $u^\mu$.

An associated subtle issue is the curl and therefore the vorticity. Covariantly one defines the curl of any four-vector $Y^\mu$ with respect to the observer $u^\mu$ by~\cite{Maartens:1998xg}
\begin{equation}
  (\textrm{curl } Y)^\mu \ := \ \eta^{\mu\nu\kappa\lambda} u_\nu h_\kappa^\sigma \nabla_\sigma Y_\lambda \,,
\end{equation}
where $h_{\mu\nu}:=g_{\mu\nu}+u_\mu u_\nu$ projects orthogonal to $u^\mu$ and the Levi-Civita alternating tensor is defined by $\eta^{0123}=-\sqrt{-g}$. (Note that our sign convention is opposite to that of Ref.~\cite{Maartens:1998xg}.) It follows that $u_\mu(\textrm{curl } Y)^\mu=0$, even if $Y^\mu u_\mu \neq 0$. If the four-vector is orthogonal, i.e., if Eq.~(\ref{eq:orthog}) holds, then the curl is particularly simple:
\begin{equation}
  (\textrm{curl } Z)^\mu \ = \ a^{-2}(0,\nabla\times\V{z}) \,.
\end{equation}
But for non-orthogonal four-vectors, the curl acquires a metric correction. The curl of the four-velocity of fluid $I$ is
\begin{equation}
  (\textrm{curl } u_I)^\mu\ =\ \frac{1}{a^2}\delta^\mu_j\epsilon^{jkl} \partial_k
    \left( v^I_l +\chi_l \right) \ = \ \frac{1}{a}(0,\betaI) \,,
    \label{eq:CurlUI}
\end{equation}
where $\betaI$ is
\begin{equation}
  \betaI \ := \ \frac{1}{a}\nabla\times\left( \V{v}_I +\Vchi \right) \,.  \label{eq:DefBetaI}
\end{equation}
This is the appropriate three-vector in the \lab frame corresponding to the fluid vorticity, which is defined by
\begin{equation}
  \omega_I^\mu \ = \ \frac{1}{2} (\textrm{curl } u_I)^\mu \,.
\end{equation}
Note that at second order, in addition to $\betaI$, products of first-order scalar perturbations appear in the spatial part of the vorticity~\cite{Matarrese:2004kq}.

\section{Electromagnetism} \label{sec:em}

We discuss how to treat electromagnetic fields in a perturbative approach. The main point is to understand the connection between the covariant four-vector representation and the three-vectors which are Fourier transformed.

The electromagnetic field is covariantly described by the Faraday tensor $F_{\mu\nu}=2\partial_{[\nu}A_{\mu]}$, where $A_\mu$ is the four-potential. Maxwell's equations are
\begin{equation}
  \nabla_{[\lambda}F_{\mu\nu]} \ = \ 0 \,,\qquad \nabla_{\nu}F^{\mu\nu} \ = \ 4\pi j^\mu \,,
\end{equation}
where the four-current is
\begin{equation}
  j^\mu \ = \ \sum_I q_I n_I u_I^\mu \,.
\end{equation}

To find the appropriate three-vector representation we split the Faraday tensor into its electric and magnetic parts relative to the observer $u^\mu$,
\begin{equation}
  F^{\mu\nu} \ = \ 2u^{[\mu}E^{\nu]} -\eta^{\mu\nu\kappa\lambda}u_\kappa B_\lambda
  \,, \qquad E^\mu u_\mu=0=B^\mu u_\mu \,.
\end{equation}
If we assume the electromagnetic field vanishes in the background, then Maxwell's equations are
\begin{eqnarray}
  \partial_i B^i & = & 0 \,,
    \\
  \partial_i (a^3 E^i) & = & 4\pi a^4 j^0 \,,
    \\
  a(\textrm{curl } a^3 B)^i & = & 4\pi a^4j^i +(a^3 E^i)' \,,
    \label{eq:MaxAmpereCov} \\
  a(\textrm{curl } a^3 E)^i & = & -(a^3 B^i)' \, .
    \label{eq:MaxInductCov}
\end{eqnarray}
We define the \lab{}electric and magnetic three-vectors as in Eq.~(\ref{eq:orthog}):
\begin{equation}
  E^\mu \ = \ {a}^{-1}( 0, \E ) \,, \qquad B^\mu \ = \ {a}^{-1}( 0, \B ) \,.
\end{equation}
The four-current may be written as
\begin{equation}
  j^\mu \ = \ {a}^{-1}( \rho_q, \V{J} ) \,,
\end{equation}
where $\V{J}$ is the \lab three-current and $\rho_q$ is the \lab charge density. For the first-order vector perturbations considered in this paper,
\begin{equation}
  \rho_q \ = \ 0 \,, \qquad \V{J} \ = \ en\left( \V{v}_p -\V{v}_p \right) \,.
\end{equation}

Maxwell's equations in the \lab{}frame then become~\cite{Subramanian:1997gi}
\begin{eqnarray}
  \nabla\cdot\B & = & 0 \,,
    \\
  \nabla\cdot (a^2\E) & = & 0 \,,
    \\
  \nabla\times(a^2\B) & = & 4\pi a^3\V{J} + \left[a^2\E\right]' \,,
    \label{eq:MaxAmpereLab} \\
  \nabla\times(a^2\E) & = & -\left[a^2\B\right]' \,.
    \label{eq:MaxInductLab}
\end{eqnarray}

Note that these results only hold in a perturbed metric if we consider the electromagnetic fields to be at the maximal order of perturbation considered and the fields to vanish in the background. In our case it makes sense to require $\E$ and $\B$ to be at first order because we are interested in magnetic field generation with vanishing initial conditions. However, one has to be very careful if one considers background electromagnetic fields and higher order perturbations, since metric perturbations would enter the expressions in Maxwell's equations.

Electromagnetic energy-momentum conservation is given by
\begin{eqnarray}
  \nabla_\nu T_{\rm em}^{\mu\nu} & = & -F^{\mu\nu}j_\nu
    \\
  & = & -F^\mu{}_\nu\sum_Iq_In_I u_I^\nu \ := \ -\sum_I K^\mu_{{\rm em},I}
    \nonumber \,.
\end{eqnarray}
For first-order vector perturbations,
\begin{equation}
  K^\mu_{{\rm em},I} \ = \ q_I n_I E^\mu \,,  \label{eq:KemI}
\end{equation}
where we dropped the second-order Lorenz-force term, $\V{v_I}\times\B$.

The curl of Amp\`ere's law, \eqref{eq:MaxAmpereLab}, provides an equation for the magnetic field with the curl of the current as its source. Using Eqs.~(\ref{eq:CurlUI}) and (\ref{eq:DefBetaI}), we find at
first order,
\begin{equation}
  (\textrm{curl } j)^\mu 
    \ = \ \frac{1}{a}\left(0,\sum_I q_In_I\betaI \right)  \,.
\end{equation}
Then the curl of \eqref{eq:MaxAmpereCov} leads to
\begin{equation}
  -\square(a^2\B) \ = \ {4\pi ena^4}\, (\betaP-\betaE) \,,  \label{eq:CurlAmpereLab}
\end{equation}
where the d'Alembert operator is defined as $\square(f):=\nabla^2f-f''$.

The above equation, together with Eqs.~(\ref{eq:PhotonVort})-(\ref{eq:ProtonVort}) which are derived from the fluids energy-momentum conservation, describe the generation and evolution of the magnetic field at first order in perturbation theory. As described in the main text, if tight coupling of electrons and photons is assumed, they can be reduced to the following evolution equation for the magnetic field:
\begin{widetext}
\begin{eqnarray}
  \left[ a^2\B - \ell^2\left(\dfrac{m_e/m_H+a_{\rm eq}/a}{a+a_{\rm eq}}\right)\square(a^2\B) \right]'
    -L^2H_0\sqrt{\Omega_{\rm rad}}\,\square(a^2\B)
    \ = \ -\dfrac{m_p}{e} a_{\rm i}^2\left(1+\dfrac{a_{\rm eq}}{a_{\rm i}}\right) \V{\beta}_{\rm i}
    \left[\dfrac{a}{a+a_{\rm eq}}\right]' \,. \label{eq:MFEvol}
\end{eqnarray}
\end{widetext}
The definition of the diffusion scales $L$, $\ell$ and a simplified version of this equation, which applies during the epoch relevant for the Harrison mechanism where $a\ll a_{\rm eq}$, are given in the main text, Eq.~(\ref{eq:Bfull}).


\end{document}